\documentstyle[aps,prl,twocolumn]{revtex}

\begin{document}

\title{
On Synchronization in a Lattice Model of Pulse-Coupled Oscillators
}

\author{
\'{A}lvaro Corral\cite{email}, Conrad J. P\'{e}rez,
Albert D\'{\i}az-Guilera, and Alex Arenas
}
\address{
Departament de F\'{\i}sica Fonamental, Facultat de
F\'{\i}sica,\\ Universitat de Barcelona, Diagonal 647, E-08028
Barcelona, Spain
}
\date{\today}

\maketitle

\begin{abstract}
We analyze the collective behavior of a lattice model of
pulse-coupled oscillators.
By means of computer simulations we find the relation between
the intrinsic dynamics of each member of the population
and their mutual interaction that ensures, in a general context,
the existence of a fully synchronized regime.
This condition turns out to be the same than the obtained
for the globally coupled population.
When the condition is not completely satisfied we find different
spatial structures.
This also gives some hints about self-organized
criticality.
\end{abstract}


     The collective behavior of large assemblies of
pulse-coupled oscillators has been investigated quite often in
the last years.  Many physical and biological systems can be
described in terms of populations of units that evolve in time
according to a certain intrinsic dynamics and interact when they
reach a threshold value \cite{HSS}.  Although it was known long
ago that the members of these systems tend to have a synchronous
temporal activity, a rigorous treatment of the problem has been
considered only in the last decade \cite{Winfree,Ermen,MS}. Up
to now, the most important efforts have been focused on systems
with long-range interactions because in this case analytical
results can be derived by applying a mean-field formalism.
Relevant is the work by Mirollo and Strogatz (MS) \cite{MS} who
discovered under which conditions mutual synchronization emerges
as the stationary configuration of the population.  Later, the
study has been generalized to other situations
\cite{Kura,Chiu,CA,Abbpre,TMS,EPG}.

     When the oscillators form a finite dimensional lattice
where only short-range interactions are allowed, the spectrum of
behaviors is more complex. For instance, some recent papers have
related lattice models of pulse-coupled oscillators to models
displaying self-organized criticality (SOC)
\cite{tesikim,PRL,MT,Bottani,Herz}, systems which self-organize, due
to its own dynamics, into a critical state with no
characteristic time or length scales
\cite{BTW}. Of particular interest for us is the model proposed
by Feder and Feder (FF) \cite{FF} to study stick-slip processes
in earthquake dynamics. In such model each cell of a 2d square
lattice is described physically in terms of a state variable
$E(t)$ (hereafter called energy) that evolves linearly with
time. Once the energy of a cell reaches a threshold value
($E_{i,j} \geq E_c=1$) it becomes critical and "fires"
transferring energy to its nearest neighboring cells according
to the following rules
\begin{equation}
\begin{array}{l}
E_{nn} \rightarrow  E_{nn} +  \varepsilon   \\
E_{i,j} \rightarrow  0,
\end{array}
\label{ff}
\end{equation}
where $ \varepsilon $ is the strength of the coupling. In turn,
some of these neighboring cells may become critical generating
an avalanche that propagates through the lattice. When one
avalanche is triggered the intrinsic dynamics is stopped and
only when it is over does the driving act again. In this way,
there is a clear separation of two time scales. By identifying
the state variable $E$ with a voltagelike magnitude one can
establish an analogy between the FF model and some models of
integrate-and-fire oscillators.

     Without any other ingredient the FF model displays, in the
stationary state, and for open boundary conditions, relaxation
oscillations (RO) which give rise to spatial structures formed
by large assemblies of units, all with the same phase. In this
sense we could talk about a macroscopic (local) degree of
synchronization. However, when a dynamical noise is added to the
system new collective behaviors appear, since the synchronized
state is no longer an attractor of the dynamics. For instance,
for $\varepsilon=0.25$
\cite{tesikim}, the distribution of avalanche sizes follows a
power-law decay characteristic of SOC.

     In the short range models described above the energy of
each oscillator varies with a constant driving rate $f(E) \equiv
\frac{dE}{dt}=C$. The analysis has been extended to linear
driving rates \cite{PRL} finding new conditions to observe RO
even in presence of noise. It would be interesting to consider a
more general context, where $f'(E)$ can take an arbitrary shape
with the only restriction, $f(E)>0$.  There is another
underlying hypothesis in (\ref{ff}) that restricts the range of
interest of these models: the strength of the coupling
$\varepsilon$ is constant. In many studies of biological
pacemakers it is assumed that the response of a cell due to an
external stimulus is a function of the phase, $\Delta(\phi)$,
which depends on the current state of the oscillators
\cite{KI,Torras}. The phase-shift induces an
energy-shift $\varepsilon(E)$ (hereafter called the energy
response curve (ERC)) that, in general, is a non-constant
function of $E$. Our goal is to investigate a wide variety of
models with arbitrary ERC and driving rate $f(E)$ beyond those
considered in previous works. Although in such a general
situation a broad range of different regimes can be observed, we
will focus our study on the conditions required for the system
to develop a fully synchronous stationary state, emphasizing the
relevance of the response of a given oscillator at its reset
point, $\varepsilon(E=0)$.
The analysis also provides information about the
possible nature of SOC.  Our interest mainly concerns coupled
map lattice models, but to start the discussion a mean-field
model is introduced to clarify several dynamic aspects.

     Let us consider a system of globally coupled oscillators
which interact through a given $\bar{\varepsilon}(E)$.  The
dynamics of each unit is given by
\begin{eqnarray}
\frac{dE_i}{dt} &=& f(E_i) + \bar{\varepsilon}(E_i) \delta (t -t_j)
\label{edete}
\end{eqnarray}
plus the reset condition for $E_i \geq 1$. Here $t_j$ denotes
the time at which unit or group $j$ fires.
In
this description both time scales (driving and coupling) appear
in the same equation. We also could consider another equivalent
description where both time scales are separated such as in (1),
but this fact implies to substitute $\bar{\varepsilon}(E)$ by
$\varepsilon(E)$ related through
\begin{equation}
\int_{E}^{E+\varepsilon(E)} \frac{dE'}{\bar{\varepsilon}(E')}=
\int_{t_j^-}^{t_j^+}\delta(t-t_j) dt =1.
\label{ecintegral}
\end{equation}
To go from $\bar{\varepsilon}(E)$ to $\varepsilon(E)$ is
trivial, but the inverse implies to deal with an integral
equation that, in general, can only be solved numerically. Now,
it is simple to derive sufficient conditions to ensure perfect
entrainment between both oscillators. The method consists of
applying the following transformation, used in different
contexts by several authors \cite{Abbpre,PRE},
\begin{equation}
y=\varepsilon_0 \int_{0}^{E} \frac{dE}{\bar{\varepsilon}(E)},
\label{ydee}
\end{equation}
where $\varepsilon_{0}$ is defined to ensure that $y(1)=1$.
By substituting
into (\ref{edete}) we get
\begin{equation}
\frac{dy_i}{dt}= \varepsilon_0
\frac{f(E_i)}{\bar{\varepsilon}(E_i)} +
\varepsilon_0 \delta (t - t_j).
\label{ydete}
\end{equation}
Now, the evolution of the system is described in terms of a new
variable $y$ for which the coupling is constant. A case of
particular interest is that of a zero advance at the reset point
($\varepsilon(E=0)=0$). In such case, this transformation
is well defined if the energy transferred $\forall y\ne 0$ is
constant ($\varepsilon_0$) except for $y=0$ which is exactly
zero.  This condition plays the role of a refractory time, which
provokes that the units that have fired have zero phase at the
end of the interaction process.

     A perfect analogy can be established with the problem
studied in MS \cite{MS}.  They showed that for a constant
positive coupling, no matter how small, the synchronized state
is an absorbing state of the dynamics if the driving rate is
positive and its first derivative negative. By applying these
conditions to the new function
\begin{equation}
g(y)=\varepsilon_0 \frac{f(E)}{\bar{\varepsilon} (E)}
\end{equation}
we find the following relation between the driving rate and
$\bar{\varepsilon}(E)$ that ensure the synchronization of the
population:
\begin{equation}
\frac{f(E)}{\bar{\varepsilon} (E)} > 0 \hspace{5mm} \mbox{and}
\hspace{5mm}
\frac{f'(E)}{f(E)} <
\frac{\bar{\varepsilon} '(E)}{\bar{\varepsilon} (E)}
\hspace{5mm} \forall E.
\label{conditions}
\end{equation}
This means that given the features of the driving rate it is
always possible to find the oscillators firing in unison
provided one chooses the suitable ERC.

     The interesting point is to know whether the mathematical
result we have derived above can be extended to networks with
short-range interactions.
We have considered FF coupling (\ref{ff}) with $\varepsilon(E)$ and
nonconstant driving rate with dynamical noise \cite{noise}.
It is important to
realize that in contrast with mean-field models, where
synchronization emerges in a process where clusters of
oscillators of increasing size merge with each other (absorption
process) and never break up, in a coupled map lattice model big
assemblies of oscillators with the same phase (which eventually
may break up) are generated through large RO that sweep the
whole lattice (avalanches of the size of the system). Then, for
constant, instantaneous couplings and due to the different
nature of both mechanisms, the conditions required to find
synchronization/RO are different \cite{PRL}. However, the
situation may change if one considers a non-constant ERC. To
analyze this new situation it is convenient to apply the same
arguments discussed in \cite{PRL}, but now for a state-dependent
coupling. For a square lattice with open boundary conditions,
it is not difficult to show that
a necessary condition to observe RO of the size of the system is
\begin{equation}
E(1-\phi(4 \varepsilon (0))) + 2 \varepsilon(1- \phi(4
\varepsilon (0))) \geq 1
\label{whole}
\end{equation}
where the energy and the phase $\phi$ are related through the
following expression
\begin{equation}
\phi(E)=\int_0^E \frac{dE}{f(E)}.
\label{phidee}
\end{equation}

     First of all, we observe the relevance of the response of a
cell at the reset value. In general, a non-zero advance at this
point ($\varepsilon(0) \neq 0$) gives rise to a spatial
distribution of phases after an avalanche;
it is only under these conditions that SOC can be
obtained \cite{tesikim}.  On the other hand,
it is clear that if $\varepsilon(0) = 0$ the inequality
(\ref{whole}) is always satisfied, for any driving rate and ERC.
This means that the appearance of RO involving all the sites
implies a perfect synchrony between all the elements of the
lattice, boundaries included. However, (\ref{whole}) does not
ensure the existence of those RO. It is a necessary but not a
sufficient condition for the system to achieve a complete
synchronization. We will check by simulations that condition
(\ref{conditions}) is, as in the long-range case, sufficient to
produce synchrony.

     Before giving evidence of this fact, we want to mention
some conclusions that can be extracted from this result.
Although, in general, one cannot map straightforwardly
results from mean-field theories to coupled map lattices models,
for these particular systems and as long as synchronization is
concerned, conditions (\ref{conditions}) which are strictly
derived in a mean-field frame, can be applied to short range
systems.  This assumption is also in agreement with a conjecture
proposed in \cite{MS} for the special case of $f'(E)<0$ and
$\bar{\varepsilon}'(E)=0$.

     Several models have been considered in our simulations
finding a complete agreement with the theory. In particular, we
have performed simulations for the Peskin's model \cite{Peskin},
where the driving rate is given by
\begin{equation}
f(E)=\gamma\left( K - E \right),
\label{peskin}
\end{equation}
where $\gamma$ gives the slope of the driving rate and $ K =
\left( 1-e^{-\gamma} \right) ^{-1}$. Starting from a random
distribution of phases and for $\gamma > 0$ ($f'(E)<0$) we have
observed that the population always synchronizes when
$\bar{\varepsilon}(E)$ is a positive function of $E$ with
$\bar{\varepsilon}'(E) \geq 0$, provided that
$\varepsilon(0)=0$.
However, for $\gamma = 0$
a monotonously increasing $\bar{\varepsilon}(E)$ is needed.

\begin{figure}
\caption[]{Number of avalanches (filled) and time in period units
(hollow) that a $32 \times 32$ lattice needs to get a complete
synchronization as a function of $\gamma - \gamma'$ in a $\log -
\log$ scale.
The symbols are averages over 25 realizations and the error bars
(which for the avalanches are of the size of the symbol)
correspond to the standard deviation.
The straight line shows the $(\gamma-\gamma')^{-1}$ behavior.}
\label{32x32}
\end{figure}

     To complete this idea we have studied the time required for
the population to synchronize, starting with random phases between
$0$ and $1$, for an ERC given by
\begin{equation}
\bar{\varepsilon}(E)=\varepsilon_0\gamma' (K'-E),
\label{peskinerc}
\end{equation}
where $ K' = \left( 1-e^{-\gamma'} \right) ^{-1}$, in a $32
\times 32$ lattice.
We have plotted our results in Fig.~\ref{32x32} for open
boundary conditions.
There we can see
how this time and the number of avalanches diverge as
the difference between $\gamma$ and $\gamma'$
approaches zero.
We have also checked a driving rate and an ERC both given
by power laws and the results are very similar; the time
required to synchronize the system grows as
$(\alpha-\alpha')^{-1} $, where $ \alpha $
and $ \alpha '$ are the exponents
of the driving rate and of the ERC, respectively.
Finally we have simulated a driving rate of the type
(\ref{peskin}) with a power-law ERC. For an exponent
$ \alpha ' \geq 1$ the
system synchronizes for any value of $\gamma$; in particular the
time needed to obtain the fully synchronized state grows
exponentially as $e^{-\gamma/2}$ when $ \alpha' =1$.
On the other hand, for $ \alpha '<1$
there is only a range of values of $\gamma$ for which we get
synchronization.

     All these results are in agreement with our prediction
(\ref{conditions}). At this point it is important to remark
that when the driving and the ERC have the same
functional dependence on $E$ the inequality can be satisfied for
all
$E$ and the divergence of the time appears when both curves
$f'/f$ and $\bar{\varepsilon}'/\bar{\varepsilon}$ approach each other.
However when the dependence is different both curves can cross
and this leads to a more complex behavior. Finally, the case of
a power-law ERC with $\alpha' \geq 1$ is very important because in
this case the transformation (\ref{ydee}) cannot be performed
and therefore no mapping with the MS result can be done;
nevertheless (\ref{conditions}) still represents the sufficient
condition to get a fully synchronized regime.
Lattices with periodic boundary conditions have also
been checked and the
conclusions are the same than for open ones; furthermore, the
time required to synchronize scales in the same way.

  We have also investigated the behavior of the model when one does
not expect synchronization for a constant ERC, provided that
$\varepsilon(E=0)=0$.
First of all, let us recall the behavior of two coupled oscillators
\cite{MS}.
The phase of the one oscillator when the other one arrives to
the threshold
transforms according to
\begin{equation}
\phi_0 \rightarrow  1-\phi\left(E(\phi_0)+\varepsilon \right)
\label{noseque}
\end{equation}
provided that $E(\phi_0)+\varepsilon < 1$,
otherwise the oscillators synchronize.
This transformation has always at least one fixed point,
$\phi_{0}^{*}$,
which leads to different behaviors depending on the slope
of $f(E)$. Thus, for $f'(E)<0 \hspace{1em}(\forall E)$,
$\phi_{0}^{*}$
is unique and unstable and the two oscillators will always
synchronize for any positive value of $\varepsilon$ \cite{MS}.
However, for $f'(E)>0 \hspace{1em}(\forall E)$ the stability of
the fixed point changes,
and $\phi_{0}^{*}$ becomes an attractor,
which means that the oscillators can either be phase-locked, when
$\varepsilon < 1-E(\phi_0)$, or synchronized, otherwise.
For the Peskin's model (\ref{peskin}) $\phi_{0}^{*}$ corresponds to
\begin{equation}
\phi_0^{*} = 1/2-1/\gamma \sinh^{-1}\left(\varepsilon
\sinh(\gamma/2)\right).
\label{noseque2}
\end{equation}

     This result enables us to perform a qualitative analysis of
the lattice model with nearest-neighbor interactions. Starting
from a random distribution of phases, and only for periodic
boundary conditions, the oscillators will tend to be locally
synchronized, or phase-locked depending on their phases
and the parameters' values.
This makes us to suspect the existence
of well defined spatiotemporal structures of phase-locked
oscillators in the stationary state, of the same kind as those
described in \cite{chinchon}, for which simple return maps can
be written.  The most simple of these configurations one can
imagine is a chessboard lattice where "black" sites have the
same value of the phase ($\phi_0$) when all the "white" sites
arrive to the threshold. Once the white ones have fired, the
black ones are driven up to the threshold value and one gets the
same structure as before with a phase $\phi_0$ that transforms
according to (\ref{noseque}) replacing $ \varepsilon $ by $4
\varepsilon $. This means that there exists a fixed point for
this structure, whose value is the same than that obtained from
(\ref{noseque2}) replacing $\varepsilon$ by $4 \varepsilon$,
that is an attractor for the dynamics of the lattice.  It is
important to remark that this structure is characterized by the
fact that a given oscillator is phase-locked with its four
neighbors.  Other simple structures arise when an oscillator can
be synchronized with one neighbor and phase-locked with the
remaining ones (all these neighbors with the same phase), and so
on, as far as the number of synchronized neighbors is constant
through the lattice.  What we have observed in our simulations
is that these structures are attractors of the dynamics within
different domains of the space of initial conditions.
The corresponding fixed points are given by
(\ref{noseque2}) changing $\varepsilon$ by $4 \varepsilon$,
$3 \varepsilon$, or $2 \varepsilon$.  Nevertheless, these
structures are not the only attractors since more complicated
patterns involving more than two different phases also can be
built.  Among of all these configurations, the "chessboard" one
is the most relevant because it has the larger domain of
attraction, although the relative weight depends on the
parameters as well as on the system size, and it is the most
stable one in the sense that small fluctuations break the other
structures in its favor \cite{futuro}.  We want to
point out that these phase-locked states are characteristic of
lattice models with short-range interactions, since they have no
counterpart with models of all-to-all pulse-coupled oscillators.

     Finally, we believe that the tendency of our model either to
synchronize or to form phase-locked structures allows us
to go one step further in the current understanding of SOC
phenomena. Middleton and Tang \cite{MT} noticed that SOC
appears, in a uniformly driven model and a slightly different
coupling, as a consequence of marginally stable phase-locking
between neighbors.  We have shown how this marginal stability
(which corresponds to the equality in the r.h.s. of
(\ref{conditions})) is broken in favor of a synchronization or a
stable phase-locking depending on the driving rate and on the
ERC. Thus, although further investigation along this line would
be required, one can conjecture that SOC is critical in the
sense that balances the tendency into one or another direction.

     In this paper we have shown how is possible to reduce a
general population of biological oscillators to a simple model
with constant ERC, that allows analytical results for all-to-all
coupling by means of a very easy
transformation. Surprisingly, short-range simulations verify the
same conditions for the synchronization that the long-range
version of the model. Moreover, we also find new states of the
system with no analogous in the long-range case. These behaviors
allow us to give some hints about the origin of SOC.

     The authors are indebted to L.F. Abbott, K. Christensen, A.
Herz and C. Vanvreeswijk for very fruitful discussions and to S.
Bottani for sending us a copy of \cite{Bottani} prior to
publication. This work has been supported by CICyT of the
Spanish Government, grant \#PB92-0863.


\begin{references}
\bibitem[*]{email}e-mail address: {\tt alvaro@ulyses.ffn.ub.es}

\bibitem{HSS}
J.J. Hopfield, Phys. Today {\bf 47}, 40 (1994);
S.H. Strogatz and I. Stewart, Sci. Am. {\bf 269} (No. 6), 68 (1993).

\bibitem{Winfree}
A.T. Winfree, {\em The Geometry of Biological Time}
(Springer-Verlag, New York, 1980).

\bibitem{Ermen}
G.B. Ermentrout, J. Math. Biol. {\bf 22}, 1 (1985).

\bibitem{MS}
R.E. Mirollo and S.H. Strogatz,
SIAM J. Appl. Math.  {\bf 50}, 1645 (1990).

\bibitem{Kura}
Y. Kuramoto, Physica {\bf 50D}, 15 (1991).

\bibitem{Chiu}
C.C. Chen,
Phys. Rev. E {\bf49}, 2668 (1994).

\bibitem{CA}
C. Vanvreeswijk and L.F. Abbott,
SIAM J. Appl. Math.  {\bf 53}, 253 (1993).

\bibitem{Abbpre}
C. Vanvreeswijk and L.F. Abbott,
Phys. Rev. E {\bf 48}, 1483 (1993).

\bibitem{TMS}
M. Tsodyks, I. Mitkov, and H. Sompolinsky,
Phys. Rev. Lett. {\bf 71}, 1280 (1993).

\bibitem{EPG}
U. Ernst, K. Pawelzik, and T. Geisel,
Phys. Rev. Lett. {\bf 74}, 1570 (1995).

\bibitem{tesikim}
K. Christensen, Ph.D. thesis, University of Aarhus, Denmark, 1992.

\bibitem{PRL}
A. Corral, C.J. P\'{e}rez, A. D\'{\i}az-Guilera, and A. Arenas,
Phys. Rev. Lett. {\bf 74}, 118 (1995).

\bibitem{MT}
A.A. Middleton and C. Tang,
Phys. Rev. Lett. {\bf 74}, 742 (1995).

\bibitem{Bottani}
S. Bottani, Phys. Rev. Lett. {\bf 74}, 4189 (1995).

\bibitem{Herz}
A.V.M. Herz and J.J. Hopfield,
Phys. Rev. Lett. {\bf 75}, 1222 (1995);
J.J. Hopfield  and A. V. M. Herz,
Proc. Natl. Am. Soc. USA {\bf 92}, 6655 (1995).

\bibitem{BTW}
P. Bak, C. Tang, and K. Wiesenfeld,
Phys. Rev. A {\bf 38}, 364  (1988).

\bibitem{FF}
H.J.S. Feder and J. Feder,
Phys. Rev. Lett.  {\bf 66}, 2669 (1991).

\bibitem{KI}
J.P. Keener, F.L. Hoppensteadt, and J. Rinzel,
SIAM J. Appl. Math {\bf 41}, 503 (1981);
N. Ikeda, Biol. Cybern. {\bf 43}, 157 (1982).

\bibitem{Torras}
C. Torras, J. Math. Biol. {\bf 24}, 291 (1986);
J.P. Segundo and A.F. Kohn, Biol. Cybern. {\bf 40}, 113 (1981).

\bibitem{PRE}
W. Gerstner, Phys. Rev. E {\bf 51}, 738 (1995).

\bibitem{noise}
The dynamical noise simply chooses at random one oscillator
to start the avalanche
between those that arrive simultaneously to the threshold.

\bibitem{Peskin}
C.S. Peskin, {\em Mathematical Aspects of Heart Physiology},
Courant Institute of Mathematical Sciences (New
York University, New York, 1975).

\bibitem{chinchon}
Q. Zhilin, H. Gang, M. Benkun, and T. Gang,
Phys. Rev. E {\bf 50}, 163 (1994).

\bibitem{futuro}
A. Corral, C.J. P\'{e}rez, A. D\'{\i}az-Guilera, and A. Arenas,
(in preparation).

\end{references}
\end{document}